# SISTEM INFORMASI AKADEMIK ONLINE SEBAGAI PENUNJANG SISTEM PERKULIAHAN

Chairil Anwar [1], Spits Warnars H.L.H. [2]

[1] Fakultas Teknologi Informasi, Universitas Budi Luhur
Jl. Petukangan Utara, Kebayoran Lama, Jakarta Selatan 12260, Indonesia
[2] Department of Computing and Mathematics, Manchester Metropolitan University
All Saints, Manchester M15 6BH, United Kingdom
[1] ch_4nw4r@yahoo.com, [2] s.warnars@mmu.ac.uk

**ABSTRACT**

The development of Internet technology give a chance for application in all kind of humans activity include education specially for higher education in university in order to increase education quality. In this paper we will discuss about the importance of the elements which must be fulfilled when we build the online academic application with web technology which will help teaching process in university. Also we will discuss to implement open source software as foundation in order to deploy the information system with low cost without losing the performance. In the future assembling internet technology on education and training will be need in order to enhancement and distribution education quality, specially in Indonesia where has many islands separate with sea and mountain. Students and faculty will enjoy access to multiple resources from one location, delivering search results with a range of content including encyclopedia articles, multimedia, related Web sites, magazines, and much more. Enables schools to implement an online academic program. It equips schools to host online courses, conduct live and blog-style communication between faculty and students, track grades, automate test scoring, store course materials, and much more.

**KEY WORDS**
Akademik Online, persaingan, Sistem Informasi Akademik, Sistem Perkuliahan, e-learning,

**1. Pendahuluan**

Sistem Informasi adalah pengaturan orang, data, proses dan teknologi informasi yang saling berinteraksi untuk mengumpulkan, memproses, menyimpan dan menyediakan data sebagai sebuah informasi/keluaran yang dibutuhkan untuk mendukung kegiatan sebuah organisasi [9]. Sistem informasi didalam sebuah organisasi brtugas untuk menangkap dan mengelola data untuk menghasilkan informasi yang berguna dan efektif yang mendukung kegiatan organisasi dan seluruh level manajemen yang menggunakan, konsumen, suplier dan rekanan bisnis Sistem informasi akan membutuhkan dukungan teknologi informasi seperti mana yang biasanya sudah lasim bahwa sistem information tidak akan berarti apa-apa tanpa adanya dukungan teknologi informasi.

Banyak organisasi baik organisasi pemerintah atau swasta yang menggunakan sistem informasi sebagai sebuah alat untuk bersaing dalam persaingan bisnis atau pelayan umum untuk organisasi pemerintah. Dalam rangka menggunakan sistem informasi '5 forces competition' Michael Porter dapat diterapkan dalam persaingan [9][10]. Organisasi harus memberikan perhatiannya kepada:
a. pesaing, yang melakukan kegiatan usaha yang sama atau organisasi yang sama.
b. Kemungkinan adanya pendatang baru dalam bisnis yang akan menjadi pesaing
c. Kemungkian adanya pengganti yang kemungkinan juga akan menjadi pesaing.
d. Konsumen, sebagai yang menggunakan produk
e. Suplier yang menyediakan kebutuhkan proses bisnis dalam organisasi

Jika dipetakan dengan '5 forces competition' Michael Porter maka kira-kira sebuah perguruan tinggi sebagai sebuah organisasi yang menjalankan sistem informasi akademik akan mempunyai peta berikut ini:

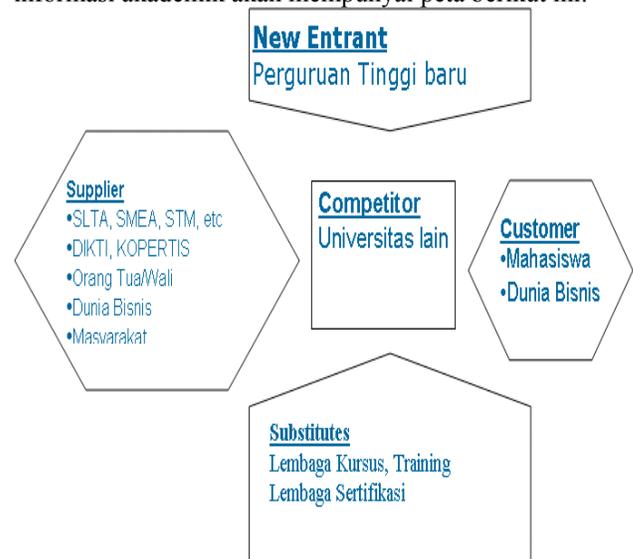

Gambar 1. 5 forces competition untuk perguruan tinggi

Dengan memetakan seluruh entitas yang berhubungan dengan sistem informasi akademik untuk sebuah perguruan tinggi,diharapkan perguruan tinggi sebisa mungkin mengelola kelima entitas ini, sehingga tujuan organisasi yang tertuang dalam visi dan misi dapat tercapai.

## 2. Teknologi Web

Teknologi web saat ini mengalami kemajuan yang sangat pesat.Perkembangan pesat ini diawali dengan dibuatnya protokol (Hypertex Transfer Protokol (HTTP), pada awalnya HTTP ini hanya untuk menampilkan dokumen text, dimana dalam dokumen tersebut tersebut ada link ke dokumen lain , baik terletak diserver yang sama atau di server yang jauh.Semua ini pada awalnya masih berbasis teks,kemudian ditemukanlah browser yang berbasis grafis.Diciptakannya browser berbasis grafis menambah kemampuan HTTP yaitu penambahan gambar pada dokumen,kemudian suara dan video pun dapat dikirim,didengar dan dapat dilihat secara langsung,semuanya itu berada diatas protokol HTTP. Perkembangan teknologi web tersebut juga diikuti berkembangnya basis data pada web,menjadikan teknologi ini semakin siap memberikan alternatif solusi permasalahan yang dihadapi manusia.

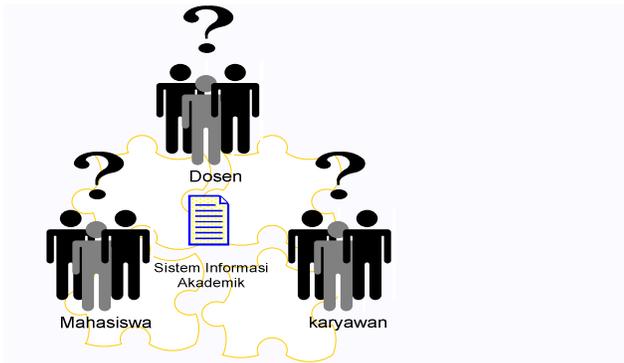

Gambar 2: Rumitnya Manual Sistem Informasi Akademik

Dalam melayani terlaksananya sistem belajar mengajar maka sistem pelayanan akademik sangat dibutuhkan dan merupakan sebuah hal yang sangat vital untuk terlaksananya proses belajar mengajar secara kontinuitas dan berkualitas. Tanpa adanya sebuah sistem informasi akademik maka proses belajar mengajar akan menjadi bukan sebuah proses belajar mengajar. Dalam dunia pendidikan tak kalah pentingnya sebagai contoh pengisian kartu rencana studi  merupakan salah satu komponen syarat dalam pendidikan.Pengisian kartu rencana studi yang dilakukan secara manual dengan form atau dimasukkan oleh staff fakultas terdapat kendala yang dihadapi yaitu kendala  dari segi geografis sehingga membutuhkan mobilitas yang tinggi untuk menuju ke kampus,kurang fleksibel terhadap aktifitas yang lain,dan membutuhkan waktu yang relatif lama. Disamping itu dengan cara ini masih ada juga kendala lain yang biasa terjadi, contohnya staff salah dalam memasukkan data KRS yang diinginkan mahasiswa, alat untuk membaca form yang ditulis oleh mahasiswa salah dalam pembacaan, bisa juga kode yang ditulis di form oleh mahasiswa beda dengan yang dimaksudkan. Hal ini tentu saja akan menghambat proses belajar mengajar yang akan di selenggarakan nantinya.

Dengan adanya teknologi web dan adanya kendala-kendala tersebut maka adalah suatu ide untuk membuat aplikasi perangkat lunak yang mampu mengelola pengisian KRS  dengan baik, sehingga kendala-kendala tersebut dapat ditangani dengan perangkat lunak ini. Dengan perangkat lunak ini diharapkan dapat memberikan alternatif lain terhadap sistem pengisian KRS yang ada sekarang ini. Sehingga masalah-masalah yang timbul pada pelaksanaan KRS secara manual dapat teratasi. Sehingga diharapkan pelaksanaan sistem informasi yang berbasiskan teknologi informasi dapat terlaksana, terlebih untuk melayani mahasiswa sebagai konsumen yang harus dilayani dan terpuaskan oleh pelayanan perguruan tinggi sebagai salah satu syarat untuk bersaing dengan perguruan tinggi lain.

Gambar 3: Kartu Rencana Studi
Sumber: http://anggacoz.files.wordpress.com/2008/02

Web merupakan salah satu tekonologi internet yang telah berkembang sejak lama dan yang paling umum dipakai dalam pelaksanaan pendidikan dan pengisian KRS maupun untuk melihat nilai secara Online ( Aplikasi Akademik Online ).
Secara umum aplikasi di internet terbagi menjadi 2 jenis, yaitu sebagai berikut:

- *Synchronous System*
  Aplikasi yang berjalan secara waktu nyata dimana seluruh pemakai bisa berkomunikasi pada waktu yang sama, contohnya: *chatting*, *Video Conference*, dsb.
- *Asynchronous System*
Aplikasi yang tidak bergantung pada waktu dimana seluruh pemakai bisa mengakses ke sistem dan melakukan komunikasi antar mereka disesuaikan dengan waktunya masing-masing, contohnya: *BBS*, *e-mail*, dsb.

Dengan fasilitas jaringan yang dimiliki oleh berbagai pendidikan tinggi atau institusi di Indonesia baik intranet maupun internet, sebenarnya sudah sangat mungkin untuk diterapkannya sistem pendukung Aplikasi Akademik Online berbasis Web dengan menggunakan sistem *synchronous* atau *asynchronous*, namun pada dasarnya kedua sistem diatas biasanya digabungkan untuk menghasilkan suatu sistem yang efektif karena masing-masing memiliki kelebihan dan kekurangannya. Dibeberapa negara yang sudah maju dengan kondisi infrastruktur jaringan kecepatan tinggi akan sangat memungkinkan penerapan teknologi multimedia secara waktu nyata seperti *video conference* untuk kepentingan aplikasi *e-Learning*, tetapi untuk kondisi umum di Indonesia dimana infrastruktur jaringannya masih relatif terbatas akan mengalami hambatan dan menjadi tidak efektif. Namun demikian walaupun tanpa teknologi multimedia tersebut, sebenarnya dengan kondisi jaringan internet yang ada sekarang di Indonesia sangat memungkinkan, terutama dengan menggunakan sistem asynchronous ataupun dengan menggunakan sistem synchronous seperti chatting yang disesuaikan dengan sistem pendukung pendidikan yang akan dikembangkan.

**3. Sistem Informasi Akademik Online**

Dengan adanya sistem informasi akademik online proses pengisian KRS tidak hanya terjadi di dalam ruangan kelas saja dimana secara terpusat staff memasukkan data-data perkuliahan yang akan diambil oleh mahasiswa, tetapi dengan bantuan peralatan komputer dan jaringan, para siswa dapat secara aktif dilibatkan dalam proses pengisian KRS ini. Mereka bisa terus berkomunikasi sesamanya kapan dan dimana saja dengan cara akses ke sistem yang tersedia secara online. Sistem seperti ini tidak saja akan menambah pengetahuan seluruh siswa, akan tetapi juga akan turut membantu meringankan beban staff dalam proses belajar-mengajar, karena dalam sistem ini beberapa fungsi staff dapat diambil alih dalam suatu program komputer yang dikenal dengan istilah *agent*. Disamping itu, hasil dari pengisian KRS ini bisa dipertanggungjawabkan terhadap pihak mana saja baik mahasiswa maupun pihak dari universitas karena proses pengisian itu sendiri langsung di lakukan oleh mahasiswa tersebut dan hasil dari pengisian itu sendiri bisa disimpan datanya di dalam bentuk *database*, yang bisa dimanfaatkan untuk mahasiswa tersebut sebagai rujukan jika sewaktu-waktu mahasiswa tersebut lupa jadwal untuk kuliah, mahasiswa dapat mengakses aplikasi akademik online tersebut sehingga aplikasi ini bisa dimanfaatkan mahasiswa juga sebagai suatu reminder dimanapun dia berada.

Gambar 4: KRS online
Sumber: http://www.sisfokampus.net/

Adapun keuntungan – keuntungan yang dapat diperoleh dari adanya sistem aplikasi akademik online ini antara lain :
- Mahasiswa terhindarkan dari kemungkinan salah mengambil matakuliah yang akan diambil, karena dari sistem aplikasi ini otomatis terhubung ke database dan mahasiswa hanya perlu melihat nama matakuliah yang mereka ambil, adapun kode matakuliah sendiri ada di sampingnya.
- Proses pengisian KRS ini sudah diantisipasi oleh sistem, misalnya ada mahasiswa yang akan mengambil satu matakuliah dan mahasiswa tersebut belum mengambil matakuliah prasyaratnya, maka sistem akan menolak untuk memasukkan matakuliah tersebut.
- Jika mahasiswa mengambil satu matakuliah dan jadwal matakuliah tersebut bentrok dengan jadwal matakuliah yang sudah dipilih sebelumnya, sistem juga akan memberikan warning dan menolak untuk memasukkan matakuliah tersebut.
- Jika satu matakuliah sudah mencapai kapasitas maksimum yang telah ditentukan maka mahasiswa tidak akan bisa mengambil matakuliah tersebut, hal ini perlu dilakukan agar proses belajar mengajar di dalam kelas menjadi nyaman, sehingga mahasiswa tidak perlu berdesak-desakan dalam mengikuti suatu matakuliah.
- Dengan adanya sistem aplikasi akademik ini, jadwal pengisian KRS serta jadwal pembayaran untuk semester / periode tersebut dapat diatur, sehingga memudahkan staff universitas dalam melihat data-data pengisian krs dan juga data-data pembayaran keuangan mahasiswa.

- Staff Universitas dapat melihat jumlah peminat dari satu matakuliah, sehingga bisa memutuskan apakah matakuliah tersebut akan diselenggarakan atau tidak dalam perkuliahan. Jika jumlah mahasiswa yang mengambil matakuliah tersebut memenuhi syarat, maka bisa langsung diambil keputusan bahwa perkuliahan tersebut akan diselenggarakan, begitu juga sebaliknya jika satu matakuliah kurang diminati oleh mahasiswa dan jumlah pesertanya hanya sedikit, staff dapat segera mengumumkan kepada mahasiswa bahwa perkuliahan untuk matakuliah tersebut tidak jadi diselenggarakan untuk periode tersebut.

### 4. Konfigurasi Sistem

Gambar berikut menunjukkan struktur global dari sistem pendukung untuk *e-Learning*. Pemakai sistem dalam hal ini mahasiswa dan staff serta dosen dapat mengakses ke sistem dengan menggunakan piranti lunak browser.

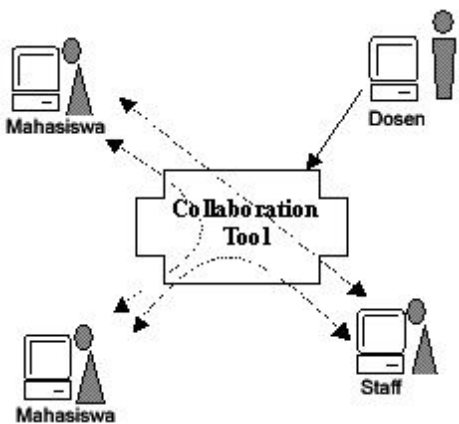

Gambar 5: Struktur Sistem

Menurut Marion A. Barfurth dalam bukunya, *Collaboration* didefinisikan sebagai kerjasama antar peserta dalam rangka mencapai tujuan bersama [1]. Collaboration tidak hanya sekedar menempatkan para peserta ke dalam kelompok-kelompok studi, tetapi diatur pula bagaimana mengkoordinasikan mereka supaya bisa bekerjasama dalam studi [2]. Saat ini penelitian di bidang kolaborasi melalui internet dikenal dengan istilah *CSCL (Computer Supported Collaborative Learning)*, dimana pada prinsipnya *CSCL* berusaha untuk mengoptimalkan pengetahuan yang dimiliki oleh para peserta dalam bentuk kerjasama dalam pemecahan masalah. Kenyataannya kolaborasi antar peserta cenderung lebih mudah dibandingkan dengan kolaborasi antara peserta dengan pengajar [5]. Gambar 5 juga menunjukkan konsep *e-Learning* dengan metoda *CSCL*, yang terdiri dari pemakai dan tool yang digunakan. Pemakai terdiri dari siswa dan pengajar yang membimbing, dimana siswa itu sendiri terbagi menjadi siswa dan siswa lain yang bertindak sebagai *collaborator* selama proses belajar.

Para peserta saling berkolaborasi dengan tool yang tersedia melalui jaringan intranet atau internet, dimana guru mengarahkan jalannya kolaborasi supaya mencapai tujuan yang diiginkan. Dalam pelaksanaan sistem *e-Learning*, kolaborasi antar siswa akan menjadi faktor yang esensial [3], terutama pada sistem *asynchronous* dimana para siswa tidak secara langsung bisa mengetahui kondisi siswa lain, sehingga seandainya terjadi masalah dalam memahami makalah yang disediakan, akan terjadi kecenderungan untuk gagal mengikutinya dikarenakan kurangnya komunikasi antar siswa, sehingga timbul kecenderungan terperangkap pada kondisi *standstill*, sehingga menyebabkan hasil yang tidak diharapkan.

Ada 5 hal essensial menurut Kedong and Jianhua [4][5] yang harus diperhatikan dalam menjalankan kolaborasi lewat internet, yaitu sebagai berikut:

a) clear, positive interdependece among students
b) regular group self-evaluation
c) interpersonal behaviors that promote each member's learning and success
d) individual accountability and personal responsibility
e) frequent use of appropriate interpersonal and small group social skills

Dalam proses kolaborasi antar siswa, pengajar / staff bisa saja terlibat didalamnya secara tidak langsung, dalam rangka membantu proses kolaborasi dengan cara memberikan arahan berupa message untuk memecahkan masalah. Sehingga diharapkan proses kolaborasi menjadi lebih lancar.

Implementasi client/server untuk sistem penunjang pendidikan berbasis kolaborasi di internet, pada dasarnya harus memiliki bagian-bagian sebagai berikut:

- *Collaboration*, untuk melakukan kerjasama antar mahasiswa dengan staff dalam pemecahan masalah yang berkaitan dengan matakuliah yang akan diambil. Kolaborasi ini bisa diwujudkan dalam bentuk diskusi atau tanya-jawab dengan memanfaatkan fasilitas internet yang umum dipakai misalnya: *e-mail, BBS, chatting*, dikembangkan sesuai dengan kebutuhan aplikasi yang akan dibuat.
- *Database*, untuk menyimpan materi pelajaran dan record-record yang berkaitan dengan proses belajar-mengajar khususnya proses kolaborasi.
- *Web Server*, merupakan bagian mengatur akses ke sistem dan mengatur tampilan yang diperlukan dalam proses pendidikan. Termasuk pula pengaturan keamanan sistem.

Pengembang aplikasi seperti ini bisa dilakukan dengan menggunakan software sebagai berikut:

| Platform OS | Linux |
|---|---|
| Web Server | Apache+Tomcat |
| Programming | Java |
| Script | Java Server Page |
| Database | MySQL / Postgres |
| Frame Work | Struts |
| Development Tool | JBuilder |

Keuntungan menggunakan *software* diatas yaitu seluruhnya merupakan *Open Source* yang bisa didownload secara gratis dari *web site* masing-masing, sehingga dalam implementasinya bisa ditekan biaya serendah mungkin, tanpa mengurangi realibilitas sistem itu sendiri. Keuntungan lainnya yaitu untuk akses ke sistem seperti ini tidak tergantung pada suatu platform operating system.
Oleh karena itu, dengan penerapan berbagai software *Open Source* seperti ini, diharapkan akan dicapai suatu sistem aplikasi yang aman, terpercaya, performance tinggi, multiplatform, dan biaya rendah.

## 5. Kesimpulan

Sejalan dengan perkembangan teknologi jaringan khususnya internet, dan pemerataan pemakaian fasilitas internet di Indonesia, maka sudah selayaknya untuk memulai penerapan teknologi ini di bidang pendidikan, yang diharapkan dapat menunjang peningkatkan mutu pendidikan khususnya pendidikan tinggi dan institusi yang relatif telah memiliki fasilitas jaringan komputer.

Dalam makalah ini telah dibahas berbagai fasilitas penunjang yang bisa dikembangkan dengan memanfaatkan teknologi internet dengan biaya yang seminimal mungkin melalui pemanfaatan *Open Source* tanpa mengurangi kualitas sistem. Faktor kolaborasi menjadi penting dalam rangka menciptakan sistem pendidikan yang lebih efektif, karena dalam sistem pendidikan sekarang ini, efisiensi waktu dan tenaga akan menjadi factor penting dalam suksesnya proses belajar mengajar.

Aplikasi Akademik Online ini sangat membantu semua pihak, baik dari pihak Universitas sendiri maupun dari mahasiswa. Dilihat dari pihak Universitas, para staff lebih ringan kerjanya, dapat langsung melihat mahasiswa yang mengisi Kartu Rencana Studi nya secara Online, dapat melihat prosentase peminat suatu matakuliah dan lain lainnya, sehingga dapat segera diambil keputusan apakah matakuliah tersebut akan diselenggarakan atau tidak.

Dilihat dari pihak mahasiswa : mahasiswa tidak perlu lagi dating ke kampus hanya untuk mengentry rencana studi, mereka bias melakukan itu dimana saja asalkan terhubung dengan internet.Mahasiswa juga lebih mudah dalam memilih matakuliah yang akan diambil, tanpa perlu takut lagi terjadi kesalahan dalam pengambilan matakuliah yang akan diambil. Selain itu, jika ternyata matakuliah yang akan diambil tersebut tidak jadi diselenggarakan, maka mahasiswa akan lebih cepat menentukan matakuliah yang lain yang akan diambil pada periode / semester tersebut.

Keberhasilan pengintegrasian system informasi akademik online ini ke Universitas sangatlah tergantung dari dukungan semua pihak yang terkait, karena dengan adanya system ini maka informasi akademik dapat dengan mudah didapat / diakses oleh staff universitas maupun mahasiswa.

Penggunaan teknologi berbasis Open Source sangat disarankan sehingga mengurangi ketergantungan terhadap suatu vendor , meningkatkan efisiensi dalam anggaran belanja TI, serta mewujudkan kampus sebagai institusi yang menghargai HAKI.